\let\includefigures=\iftrue
%
\let\useblackboard=\iftrue
%
%
\newfam\black
\input harvmac
\noblackbox
\includefigures
\message{If you do not have epsf.tex (to include figures),}
\message{change the option at the top of the tex file.}
\input epsf
\def\figin{\epsfcheck\figin}\def\figins{\epsfcheck\figins}
\def\epsfcheck{\ifx\epsfbox\UnDeFiNeD
\message{(NO epsf.tex, FIGURES WILL BE IGNORED)}
\gdef\figin##1{\vskip2in}\gdef\figins##1{\hskip.5in}
\else\message{(FIGURES WILL BE INCLUDED)}%
\gdef\figin##1{##1}\gdef\figins##1{##1}\fi}
\def\DefWarn#1{}
\def\figinsert{\goodbreak\midinsert}
\def\ifig#1#2#3{\DefWarn#1\xdef#1{fig.~\the\figno}
\writedef{#1\leftbracket fig.\noexpand~\the\figno}%
\figinsert\figin{\centerline{#3}}\medskip\centerline{\vbox{
\baselineskip12pt\advance\hsize by -1truein
\noindent\footnotefont{\bf Fig.~\the\figno:} #2}}
\bigskip\endinsert\global\advance\figno by1}
\else
\def\ifig#1#2#3{\xdef#1{fig.~\the\figno}
\writedef{#1\leftbracket fig.\noexpand~\the\figno}%
\global\advance\figno by1}
\fi
%

\def\smallfig#1#2#3{\DefWarn#1\xdef#1{fig.~\the\figno}
\writedef{#1\leftbracket fig.\noexpand~\the\figno}%
\figinsert\
in{\centerline{#3}}\medskip\centerline{\vbox{
\baselineskip12pt\advance\hsize by -1truein
\noindent\footnotefont{\bf Fig.~\the\figno:} #2}}
\endinsert\global\advance\figno by1}

\useblackboard
\message{If you do not have msbm (blackboard bold) fonts,}
\message{change the option at the top of the tex file.}
\font\blackboard=msbm10 scaled \magstep1
\font\blackboards=msbm7
\font\blackboardss=msbm5
\textfont\black=\blackboard
\scriptfont\black=\blackboards
\scriptscriptfont\black=\blackboardss

\else

\fi
%


\def\boxit#1{\vbox{\hrule\hbox{\vrule\kern8pt
\vbox{\hbox{\kern8pt}\hbox{\vbox{#1}}\hbox{\kern8pt}}
\kern8pt\vrule}\hrule}}
\def\mathboxit#1{\vbox{\hrule\hbox{\vrule\kern8pt\vbox{\kern8pt
\hbox{$\displaystyle #1$}\kern8pt}\kern8pt\vrule}\hrule}}

\def\subsubsection#1{\bigskip\noindent
{\it #1}}

\def\yboxit#1#2{\vbox{\hrule height #1 \hbox{\vrule width #1
\vbox{#2}\vrule width #1 }\hrule height #1 }}
\def\fillbox#1{\hbox to #1{\vbox to #1{\vfil}\hfil}}
\def\ybox{{\lower 1.3pt \yboxit{0.4pt}{\fillbox{8pt}}\hskip-0.2pt}}
%
%



\def\l{\left}

\def\comments#1{}



\def\II{\relax{I\kern-.10em I}}

\font\cmss=cmss10 \font\cmsss=cmss10 at 7pt
\def\IZ{\relax\ifmmode\mathchoice
{\hbox{\cmss Z\kern-.4em Z}}{\hbox{\cmss Z\kern-.4em Z}}
{\lower.9pt\hbox{\cmsss Z\kern-.4em Z}}
{\lower1.2pt\hbox{\cmsss Z\kern-.4em Z}}
\else{\cmss Z\kern-.4emZ}\fi}
\def\IR{\relax{\rm I\kern-.18em R}}
\def\IZ{\relax\ifmmode\mathchoice
{\hbox{\cmss Z\kern-.4em Z}}{\hbox{\cmss Z\kern-.4em Z}}
{\lower.9pt\hbox{\cmsss Z\kern-.4em Z}} {\lower1.2pt\hbox{\cmsss
Z\kern-.4em Z}}\else{\cmss Z\kern-.4em Z}\fi}
\def\IB{\relax{\rm I\kern-.18em B}}
\def\IC{{\relax\hbox{$\inbar\kern-.3em{\rm C}$}}}
\def\ID{\relax{\rm I\kern-.18em D}}
\def\IE{\relax{\rm I\kern-.18em E}}
\def\IF{\relax{\rm I\kern-.18em F}}
\def\IG{\relax\hbox{$\inbar\kern-.3em{\rm G}$}}
\def\IGa{\relax\hbox{${\rm I}\kern-.18em\Gamma$}}
\def\IH{\relax{\rm I\kern-.18em H}}
\def\II{\relax{\rm I\kern-.18em I}}
\def\IK{\relax{\rm I\kern-.18em K}}
\def\IP{\relax{\rm I\kern-.18em P}}

%

\def\inbar{\,\vrule height1.5ex width.4pt depth0pt}

\font\cmss=cmss10 
\def\IR{\relax{\rm I\kern-.18em R}}

%


%

\def\lp10{\ell_p^{10}}
\def\lp11{\ell_p^{11}}
\def\R11{R_{11}}

\def\frac#1#2{{#1 \over #2}}


\def\l{\left}

\def\comments#1{}




\def\cf{{\it c.f.}}
\def\M4{M_{Pl,4}}

\def\k11{\kappa_{11}}
\def\l11{\ell_{11}}
\def\tl11{\tilde{\ell}_{11}}

\def\m11{M_{11}}
\def\tm11{\tilde{M}_{11}}



\def\prl{{\it Phys. Rev. Lett.}}
\def\pr{{\it Phys. Rev.}}
\def\pl{{\it Phys. Lett.}}

\def\cqg{{\it Class. Quant. Grav.}}

\def\cmp{{\it Comm. Math. Phys.}}

\def\prs{{\it Proc. Roy. Soc.}}

\lref\lms{H. Liu, G. Moore and N. Seiberg,
``Strings in a time-dependent orbifold,''
hep-th/0204168.}
\lref\vijaysing{V. Balasubramanian, S.F. Hassan,
E. Keski-Vakkuri and A. Naqvi, ``A space-time
orbifold: a toy model for a cosmological singularity,''
hep-th/0202187.}
\lref\seibergbounce{N. Seiberg, ``From a big crunch to
a big bang: is it possible?'', hep-th/0201039.}
\lref\egks{S. Elitzur, A. Giveon, D. Kutasov and E. Rabinovici,
``From big bang to big crunch and beyond,'' hep-th/0204189.}
\lref\nappiwitten{C.R. Nappi and E. Witten,
``A closed, expanding universe in string theory,''
\pl\ {\bf B293} (1992) 309, hep-th/9206078.}
\lref\bigcrunch{J. Khoury, B.A. Ovrut, N. Seiberg, P.J. Steinhardt
and N. Turok, ``From big crunch to big bang,'' \pr\ {\bf D65}
(2002) 086007; hep-th/0108187.}
\lref\hssing{G.T. Horowitz and A.R. Steif, ``Singular
string solutions with nonsingular initial data,''
\pl\ {\bf B258} (1991) 91.}
\lref\wittenbh{E. Witten, ``On string theory
and black holes,'' \pr\ {\bf D44}\ (1991) 314.}
\lref\hsprop{G.T. Horowitz and A.R. Steif,
``Strings in strong gravitational backgrounds,''
\pr\ {\bf D42}\ (1990) 1950.} 

\lref\singtheorems{R. Penrose, \prl\ {\bf 14}\ (1965) 57;
S. Hawking, \prs\ {\bf A300}\ (1967) 182;
S. Hawking and R. Penrose, \prs\ {\bf A314}\ (1970) 589.}

\lref\value{G.T. Horowitz and R.C. Myers,
``The value of singularities,'' {\it Gen. Rel. Grav.}\
{\bf 27}\ (1995) 915; gr-qc/9503062.}
\lref\albionemil{A.E. Lawrence and E.J. Martinec,
``String field theory in curved spacetime and
the resolution of spacelike singularities,'' \cqg\
{\bf 13}\ (1996) 63; hep-th/9509149.}

\lref\squeezedgm{I learned this from
various conversations at Stanford,
and various Stanford ITP group meetings.
The apparent culprits are: A. Adams, M. Fabinger,
S. Giddings, S. Hellerman, M. Kleban, S. Kachru, N. Kaloper,
J. McGreevy, S. Shenker, E. Silverstein and L. Susskind.}
\lref\bubblebath{O. Aharony, M. Fabinger, G.T. Horowitz
and E. Silverstein, ``Clean time dependent
string backgrounds from bubble baths,'' hep-th/0204158.}

\lref\gibbons{G. Gibbons, \cmp\ {\bf 45}\ (1976) 191.}
\lref\horowitzsteif{G.T. Horowitz and A.R. Steif,
``Spacetime singularities in string theory,''
\prl\ {\bf 64}\ (1990) 260.}
\lref\simon{J. Simon, ``The geometry of null rotation
identifications,'' hep-th/0203201.}
\lref\nikita{N.A. Nekrasov, ``Milne universe,
tachyons and quantum group,'' hep-th/0203112.}
\lref\ckr{B. Craps, D. Kutasov and G. Rajesh,
``String propagation in the presence of
cosmological singularities,'' hep-th/0205101.}

\lref\shells{Y. Peleg and A. Steif, ``Phase transition for 
gravitationally collapsing dust shells in $2+1$ dimensions,''
\pr\ {\bf D51} (1995) R3992; gr-qc/9412023.}

\lref\btz{M. Ba\~nados, C. Teitelboim and J. Zanelli,
``The black hole in three dimensional spacetime,''
\prl\ {\bf 69}\ (1992) 1849, hep-th/9204099; 
M. Ba\~nados, M. Henneaux, C. Teitelboim and J. Zanelli,
``Geometry of the $2+1$ black hole,'' \pr\ {\bf D48}\ (1993) 1506,
gr-qc/9302012.}

\lref\complement{L. Susskind, L. Thorlacius and J. Uglum,
``The stretched horizon and black hole complementarity,''
\pr\ {\bf D48}\ (1993) 3743, hep-th/9306069; L. Susskind, ``String
theory and the principles of black hole complementarity,''
\prl\ {\bf 71} (1993) 2367, hep-th/9307068.}

\lref\kounnaslust{C. Kounnas and D. Lust,
``Cosmological string backgrounds from gauged WZW models'',
{\it Phys. Lett.} {\bf B289} (1992) 56, hep-th/9205046.}
\lref\cornalbacosta{L. Cornalba and M.S. Costa, ``A new
cosmological scenario in string theory'', hep-th/0203031.}
\lref\cck{L. Cornalba, M.S. Costa and C. Kounnas,
``A resolution of the cosmological singularity with orientifolds'',
hep-th/0205261.}


\Title{\vbox{\baselineskip12pt\hbox{hep-th/0205288}
\hbox{SU-ITP-02/18} \hbox{SLAC-PUB-9213}}} {\vbox{
\centerline{On the instability of 3d null singularities}}}
\smallskip
\centerline{Albion Lawrence} 
\bigskip
\bigskip
\centerline{{SLAC Theory Group, MS 81, 2575 Sand Hill Road,
Menlo Park, CA 94025, {\it and}}}
\centerline{{Department of Physics,
Stanford University, Stanford, CA 94305.}}
\bigskip
\bigskip
\noindent

String propagation on a three-dimensional Lorentzian string 
orbifold with a null singularity has been
studied by Horowitz and Steif, and more recently
by Liu, Moore and Seiberg.
We analyze the target space as a classical gravitational background.  
The singularity becomes spacelike when an arbitrarily small
amount of matter is thrown at the singularity.  This can be
seen directly by constructing a Vaidya-type
solution, or by studying the null singularity
as a limit of the $M=0$, $J=0$ BTZ black hole metric.

\medskip
\bigskip

\Date{May 2002}

\newsec{Introduction}

String theory has had great success resolving
singularities in classical supergravity which occur
at a point in space.  But
time-dependent backgrounds are poorly understood,
and the fate of dynamical or 
spacelike singularities is unknown.
Since spacelike singularities arise from reasonable initial
conditions in classical general relativity
\refs{\singtheorems}, they are an important
feature of gravitational dynamics.
Furthermore, there are good reasons to
believe that the physics of 
spacelike gravitational singularities
is qualitatively different than that
of timelike singularities (\cf\ \refs{\value}).

A handful of exactly solvable conformal field
theories corresponding to strings propagating
in time dependent backgrounds with spacelike or
null singularities have
been constructed \refs{\wittenbh,\hssing,
\nappiwitten,\nikita,\simon,
\egks,\vijaysing,\lms,\ckr}.\foot{Conformal
field theories describing string propagation in
time-dependent bacgkrounds with timelike or
no singularities have been constructed in
\refs{\kounnaslust,\cornalbacosta,\cck}.}
Despite the apparent simplicity of these CFTs,
the quantum dynamics of strings remains unclear.
The target spaces of many of 
these CFTs have closed timelike curves
\refs{\vijaysing,\egks,\ckr}, and
string propagation is singular at many
of these singularities \refs{\hsprop,\hssing}.

It is not even clear whether conformal
field theory is the correct tool for
discussing such singularities.
The dynamical singularities
demanded by the singularity theorems
signal a disease in the evolution of
classical gravitational fields.  The classical limit of
string theory -- the limit of large vevs of
quantum fields -- is not yet understood, particularly
in the closed string sector.  There are no
real principles for continuing the spacetime beyond
the singularity, nor is it obvious that
continuing past the singularity is desirable.

A related issue is that
quantum string production and 
the resulting backreaction could be large.
For example,
in the spacetimes described by
\refs{\nikita,\vijaysing},
demanding that spacetime 
quantum fields are invariant under the
orbifold group at the orbifold fixed points implies large
particle production at the singularity \refs{\squeezedgm}.
Quantum {\it string} production in time-dependent
backgrounds is still a young subject
(see \refs{\bubblebath}\ for the most complete discussion
to date),\foot{See also \refs{\cck} for a spacetime field-theoretic 
calculation of string production in a time-dependent 
background.} but preliminary results suggest that
near singularities, the backreaction from string production 
may be even more drastic than that from field theory
\refs{\albionemil}.

The purpose of this work is to argue that
in the spacetime background described in 
\refs{\hssing,\lms}, the dynamics
of a probe and the classical formation
of spacelike singularities are not separable issues.
On the one hand, the target space studied is an explicit orbifold
of string theory, so it is under control
as a conformal field theory.  It
has a null singularity and no closed timelike 
curves, and the presence of
a null Killing vector insures that there is
no particle production \refs{\gibbons}.
The system seems to be an ideal laboratory
for studying time-dependent backgrounds using conformal
field theory techniques.

On the other hand, the causal structure of the singularity
is unstable.  Any stress-energy satisfying
the positive or null energy condition, at arbitrarily
small but finite energy in three dimensions, 
will replace the singularity with
a spacelike singularity, of the kind described in
\refs{\hssing,\bigcrunch,\seibergbounce}.

We will show this first by
studying null stress energy impinging
on the singularity of \refs{\hssing,\lms}, in the classical
gravitational limit in three dimensions.  We will then argue
that this behavior is reasonable and generic
by studying the solution as a particular limit
of the BTZ black hole metric.  We will conclude with
a brief discussion of string amplitudes in
this background.

\newsec{Collapsing null matter}

The metric of the orbifold in \refs{\hssing,\lms}\ 
can be written after a suitable change of coordinates as:\foot{
These are the ``y-coordinates'' in \refs{\lms}.  The singular
submanifold at $y^+ = 0$ has a different structure
than the singularity in the original coordinates,
since the coordinate transformation is bad there.
This will not effect the calculation in this paper, 
which describes the spacetime at any point off of
the singular locus.  Of course it may be important
for the resolution of the singularity.}
\eqn\ymetric{
	ds^2 = - 2 dy^+dy^- + (y^+)^2 dy^2\ ,
}
where $y \sim y + 2\pi$ is a periodic variable.
For $y^+ < 0$, this spacetime has the structure of a 2d Lorentzian
cone times a lightlike line.  It is unclear whether the
$y^+ > 0$ region should be kept; we will not address that issue here.

To study the effects of matter impinging on the
singularity, we will send in a pulse of null stress energy,
which is circularly symmetric -- that is, translationally
invariant along $y$.  We take
the energy-momentum tensor to be:
\eqn\emomansatz{
	T_{- -} = \alpha \frac{g(y^-)}{8\pi G^{(3)}_N (- y^+)}\ ,
}
which describes null matter striking the singularity.
The dependence on $y^+$ is required for $T_{\mu\nu}$
to be conserved.
Here $G_N^{(3)}$ is the 3-dimensional Newton's constant.
We assume that $g(y^+)$ is positive definite,
and that 
\eqn\normalization{
	\int_{-\infty}^{\infty} dy^- g(y^-) = 1
}
If the circular shell is approaching from $y^+ < 0$,
the energy of the incoming shell is positive
if $\alpha > 0$.

Taking the ans\"atz for the metric:
\eqn\metricansatz{
	ds^2 = f(y^-) (dy^-)^2 - 2 dy^-dy^+ + (y^+)^2 dy^2\ ,
}
the solution to Einstein's equation is:
\eqn\solution{
	f(y^-) = -\alpha \int_{-\infty}^{y^-} dx g(x)\ .
}
Assume $g$ has compact support in the interval
$y^{-} \in [0, a]$.  For $y^- \leq 0$, the
metric is that of the null singularity in \refs{\lms}.
For $y \geq a$ the metric is:
\eqn\finalmetric{
	ds^2 = \alpha (dy^-)^2 - 2 dy^+ dy^- + (y^+)^2 dy^2\ .
}
If $\alpha > 0$, so that the energy of the incoming
shell is positive, then $y^+ = 0$ has become a 
spacelike singular locus.  $y$ describes a circle whose
circumference shrinks with the time coordinate
$y^+$.  A further coordinate transformation makes the
structure clear:
\eqn\finaltomilne{
\eqalign{
	u & = \frac{1}{\alpha} (\tau + \sigma) \cr
	v &= \tau \cr
	\phi &= \sqrt{\alpha} y\ .
}}
The metric is:
\eqn\milnemet{
	ds^2 = \frac{1}{\alpha} \left(
	- d\tau^2 + d\sigma^2 + \tau^2 d\phi^2 \right)
}
where $\psi \sim \phi + 2\pi\sqrt{\alpha}$. 
The spacelike singularity at $\tau = 0$ is of the type
discussed in \refs{\hssing,\bigcrunch,\seibergbounce,\nikita}.

This instability in the causal structure
can be understood by studying the effects
of collapsing massive dust in $2+1$ dimensions \refs{\shells}.
For a given velocity of
the shell, there is a critical value
$\mu_c$ of the mass of the shell, 
at which the metric outside the shell approaches that
of the null singularity in \refs{\hssing, \lms} as the shell shrinks
to zero size. For any mass smaller than the
critical value, the endpoint is a conical
singularity (as expected for a point mass below the
critical value $\mu_c$.) For a range of
masses larger than the critical
value, the endpoint of the collapse is the
metric \finalmetric.

\newsec{The BTZ black hole}

As noted in \refs{\simon, \lms}, the null 
singularity \ymetric\ is a
particular limit of the $M=0$, $J=0$ BTZ black hole.
This provides a nice picture of the results of the previous section.

The BTZ black hole with positive mass $M$ 
and zero angular momentum $J$ has the metric \refs{\btz}:
\eqn\btzmetric{
	ds^2 = - \left(\frac{r^2}{\ell^2} - \epsilon\right)dt^2
	+ \left(\frac{r^2}{\ell^2} - \epsilon\right)^{-1} dr^2
		+ r^2 d\phi^2\ .
}
For $\epsilon > 0$, this is a black hole with mass
\eqn\btzmass{
	M = \frac{\epsilon}{8 G_N^{(3)}}\ ,
}
in anti-de Sitter space with cosmological constant
\eqn\adscc{
	\Lambda = -\frac{1}{\ell^2}\ .
}
The horizon is at $r = \sqrt{\epsilon}\ell$, and the singularity
is at $r = 0$.  As $\ell\to\infty$ the horizon recedes to infinity.

The metric \ymetric\ can be found by taking the limit $\epsilon \to 0$
while focusing on the region near the singularity.
If $r \ll \sqrt{\epsilon}\ell$, the metric approaches
\eqn\nearsing{
	- \frac{1}{\epsilon} dr^2 + \epsilon dt^2 + r^2 d\phi^2
}
In particular, if we let
\eqn\scaling{
\eqalign{
	t & = \frac{y^+}{\epsilon} \cr
	r & = \epsilon (y^- + t) = \epsilon y^- + y^+\ ,
}}
then \nearsing\ is clearly equivalent to \ymetric.
For fixed $y^+,y^-$, this scaling limit is 
consistent so long as $\ell$ is large enough.
As $\ell \to \infty$ one recovers 3d gravity
without a cosmological constant.

The instability of the null singularity is thus
a limiting case of the fact that throwing matter
at an extreme, massless BTZ black hole with a null singularity
creates a massive black hole with a spacelike singularity
and an event horizon.  Observers placed so that they
are inside the horizon after the matter passes them, eventually
see the spacetime described in \finalmetric.
If $\ell$ is finite, a black hole solution with
a horizon exists; observers which are far enough away
will see the formation of a classically static 
black hole with an event horizon.

\newsec{Conclusions}

Our discussion has been entirely three-dimensional,
as would be natural if
the remaining seven dimensions were compactified.
If additional dimensions are not compactified
and the incoming probe is localized in these
dimensions, the solution will not be as simple.
A simpler question is, whether Planckian physics
generically becomes important to
the physics of actual higher-dimensional
probes near the singularity.\foot{This point arose during a conversation 
with T. Banks, J. Polchinski, S. Shenker, E. Silverstein
and L. Susskind; see also the conclusions 
in \refs{\lms}.  The calculation described in in
the following sentences was inspired by comments by J. Polchinski.}
For a generic classical scalar field in $d$ dimensions,
one can follow the arguments
in \refs{\hssing}, by solving the equations of
motion in the spacetime \ymetric, as is done in \refs{\lms}, 
and computing the energy-momentum tensor for
a given mode.  As in the presence of the
spacelike singularities studied in \refs{\hssing},
a classical mode of a scalar field in the spacetime \ymetric\
has diverging energy near $y^+ = 0$.
This ``instability'' is more drastic 
than the effect we have described in \S2,
where we have assumed an idealized energy-momentum
tensor with a total energy which is finite even near the singularity.

Because the null singularity is marginal,
classical probes with very small incoming three-dimensional energy
will change the causal structure of the singularity.
This change is potentially serious; for example,
once the singularity becomes spacelike there can
be large particle production near the singularity \refs{\squeezedgm}.
One might hope that a signal of this change appears
in string perturbation theory, or at least as an interesting disease
in the perturbation expansion.  It would be interesting to study 
this further, along the lines of \refs{\lms}.

However, the interpretation of worldsheet computations is
unclear once the spacelike singularity develops.  
If it is natural to continue through
the singularity at $y^+ = 0$, as advocated by \refs{\egks,\ckr}\ 
and suggested by \refs{\lms},
then one may set up scattering data
far from the singularity \refs{\ckr};  
but it is not clear if this
continuation is natural or unique.

Absent this continuation, there is no obvious S-matrix
interpretation of worldsheet correlation functions due
to the future singularity.  One may try
to keep a version of the S-matrix by keeping the cosmological
constant $1/\ell$ small but finite.\foot{I am grateful
to S. Shenker for asking about and discussing this point.}  
The $\epsilon \to 0$ limit of \btzmetric\ 
is an orbifold of anti-de Sitter space
\btz.  Correlators of vertex operators specified by their behavior
at the boundary of AdS then provide an analog of
S-matrix elements.  An incoming string will create an $M>0$ 
AdS black hole with the spacelike singularity hidden 
by a horizon from the asymptotic observer.
The nontriviality of the order of limits is consistent
with the principles of black hole complementarity \refs{\complement}.

\vskip1cm
\centerline{\bf{Acknowledgments}}
I am grateful to A. Adams, M. Fabinger, S. Hellerman, V. Hubeny, 
S. Kachru, N. Kaloper, M. Kleban,
J. McGreevy, J. Polchinski, S. Shenker, E. Silverstein and
L. Susskind for discussions.  Thanks in particular to
J.M., J.P. and E.S. for comments on the draft.
This work is supported by NSF grant PHY-9870115, by the
Stanford Institute for Theoretical Physics,
and by the DOE under contract DE-AC03-76SF00515.

\listrefs
\end